\documentclass{article}
\usepackage{authblk}
\usepackage{fullpage}
\usepackage[normalem]{ulem}
\usepackage[utf8]{inputenc}
\usepackage{graphicx}
\usepackage{url}
\usepackage{color}
\usepackage{comment}
\usepackage{soul}
\usepackage{enumitem}
\usepackage{hyperref}

\newcommand{\rubenC}[1]{\textcolor{blue}{#1}}
\newcommand{\rubenF}[1]{\footnote{\textcolor{blue}{{\sc ruben}: #1}}}

\newcommand{\celsoC}[1]{\textcolor{red}{#1}}
\newcommand{\celsoF}[1]{\footnote{\textcolor{red}{{\sc Celso}: #1}}}

\title{Network polarization, filter bubbles, and echo chambers:\\
An annotated review of measures and reduction methods}

\author{Ruben Interian\thanks{Corresponding author, Email: rinterian@id.uff.br}}
\author{Ruslán G. Marzo\thanks{Email: ruslangm@id.uff.br}}
\author{Isela Mendoza\thanks{Email: imendoza@id.uff.br}}
\author{Celso C. Ribeiro\thanks{Email: celso@ic.uff.br}}

\affil{Institute of Computing, Universidade Federal Fluminense,\\ Niterói, RJ 24210-346, Brazil}

\date{}

\begin{document}

\maketitle

\begin{abstract}
    Polarization arises when the underlying network connecting the members of a community or society 
    becomes characterized by highly connected groups with weak inter-group connectivity. The increasing polarization, the strengthening of echo chambers, and the isolation caused by information filters in social networks are increasingly attracting the attention of 
    researchers from different areas of knowledge such as computer science, economics, social and political sciences. 
    This work presents an annotated review of network polarization measures and models used to handle the polarization. 
    Several approaches for measuring polarization in graphs and networks were identified, including those based on homophily, modularity, random walks, and balance theory. 
    The strategies used for reducing polarization include methods that propose edge or node editions (including insertions or deletions, as well as edge weight modifications), 
    changes in social network design, or changes in the recommendation systems embedded in these networks. 
    \\ 
    
\end{abstract}


\section{Introduction}

Polarization is a well-known phenomenon  increasingly attracting the attention of the media, politicians, influencers, and researchers. According to the Oxford dictionary, \textit{polarization is the division of a group} (or a society, or a network) \textit{into sharply contrasting subgroups, communities, or sets of opinions or beliefs}~\cite{misc_polarization}. 

Some degree of polarization is unavoidable in any democratic system. 
The excess of political homogeneity may eliminate the presence of democratic alternatives~\cite{2003Sunstein}. 
However, extreme polarization can lead to gridlocks or even violent conflicts~\cite{2015Garcia}. 

In his farewell address back in 1796, George Washington predicted that factions, or monolithic parties, would yield political sectarianism~\cite{1999Washington}. 
In the 19th century, John Stuart Mill claimed that the dialogue across lines of political difference is a key prerequisite for sustaining a democratic citizenry \cite{1859Mill}. 
Hannah Arendt asseverated that debate is irreplaceable for forming enlightened opinions that reach beyond the limits of one's own subjectivity to incorporate the standpoints of others \cite{1968Arendt}.
World leaders have often expressed concern about raising social and political polarization~\cite{2018UN}.
From sociologists to economists, from politicians to the media, many are interested in studying the behavior and interactions in social networks that rule the opinion formation process. 

Two of the main factors that shape people's opinions are confirmation bias and social influence~\cite{2021Liu}. According to Vicario et al.~\cite{2017DelVicario}, the observed polarization of offline and online communities might be the result of the conjugate effect of these two forces. 

Confirmation bias is the tendency to process information by seeking or interpreting only those facts consistent with one's existing beliefs~\cite{misc_eb}. In short, confirmation bias tends to favor information people are already convinced of. 
Even though this phenomenon may be largely unintentional, it often results in completely ignoring part of the existing information, causing significantly less contact with contradicting viewpoints. This isolation caused by confirmation bias and other information filters, such as content recommendation systems, is called by the term \textit{filter bubble}. 
Unlike confirmation bias, social influence is the process under which one's opinions or behaviors are actually affected by others~\cite{Gass2015,2017DelVicario}. 

Polarization is characterized by an increasing intra-group agreement (i.e., between individuals with similar beliefs), while at the same time, there is a deepening inter-group disagreement (i.e., between individuals identified with groups with contrary beliefs)~\cite{2008Buskens}. Social networks and mass media are places where this phenomenon manifests itself in a strong way~\cite{2018Interian}. 
However, polarization and hostility are increasingly shifting from social media to the real world, as it was demonstrated by several political events, such as the protests of the Yellow Vest movement~\cite{misc_yellow_vests} in France in 2018, the protests after George Floyd's death~\cite{misc_george_floyd} in 2020, the US Capitol attack~\cite{misc_capitol} in 2021, and the convoy protests~\cite{misc_freedom_convoy} in Canada in 2022. 

Studies have investigated the echo chamber effect on information spreading, showing that some groups can transmit information, on average, to a larger audience than others~\cite{Cota2019}. 
Polarized groups are often related to the increased spreading of fake news. 
However, polarization and fake news spreading are different processes. Polarization is about strengthening and isolating groups or communities, while fake news propagation reflects specific information diffusion within these groups. 
In addition, fake news may propagate even in the absence of polarization (under the form of rumors or misinformation), although high polarization levels facilitate their spreading. 

Community detection methods~\cite{2006NewmanA,2004Newman} are closely related to polarization detection and measurement. Studies and reviews about community detection methods appeared, e.g., in~\cite{2019ElMoussaoui,2016Yang}. 
However, community and polarization detection should not be confused. 
Community detection amounts to the identification of membership in groups. 
On the other hand, when detecting or measuring polarization, the attributes reflecting group membership are generally already known or estimated, and one seeks to identify the strength of inter-group and intra-group connections. 

In the literature, groups formed around a shared narrative are frequently called echo chambers. As defined by Cinelli et al.~\cite{2021Cinelli}, echo chambers are environments in which the opinions or beliefs of people about some topic are reinforced due to repeated interactions with peers or sources having similar tendencies and attitudes. 
Some social networks show a massive presence of echo chambers, while in others, their presence is reduced~\cite{DeFrancisciMorales2021}. The terms \textit{group}~\cite{2009Currarini}, \textit{community}~\cite{2006NewmanA}, \textit{gated community}~\cite{1997Turow}, 
\textit{filter bubble}~\cite{2017Spohr}, 
and \textit{echo chamber}~\cite{2021Cinelli}, have different shades of meaning, but are often used as synonyms in the literature. 

Articles about different topics on the general subject of polarization appear in journals from different areas of knowledge, such as computer science~\cite{2016Garimella_Quantifying,2021IntMorRib}, economics~\cite{2009Currarini,2018Kawada}, or social and political sciences~\cite{2016Flaxman,2010maoz}. 
Researchers refer to phenomena related to polarization using different terms such as \textit{controversy}~\cite{2016Garimella_Quantifying}, \textit{disagreement}~\cite{2021Chen}, \textit{conflict}~\cite{2017Ciampaglia}, and even 
\textit{cyberbalkanization}~\cite{2014Bozdag}, in addition to polarization itself. 
We will mostly treat them as synonymous, unless otherwise stated. 

Polarization manifests mainly in mass media and social, interaction, and collaboration networks. In this review, we are interested in exploring \textit{network polarization} specifically.
It is defined as a phenomenon in which the underlying network connecting the members of a society or community is composed of highly connected groups with weak inter-group connectivity~\cite{2012Conover}. 
The polarization of posts or users of a social network may also be assessed independently, without considering the underlying graph or network structure. 
In this case, the profile of each post or user is evaluated, but the connections between them are not used. 

\begin{figure}[t]
\centering
\includegraphics[width=0.59\textwidth]{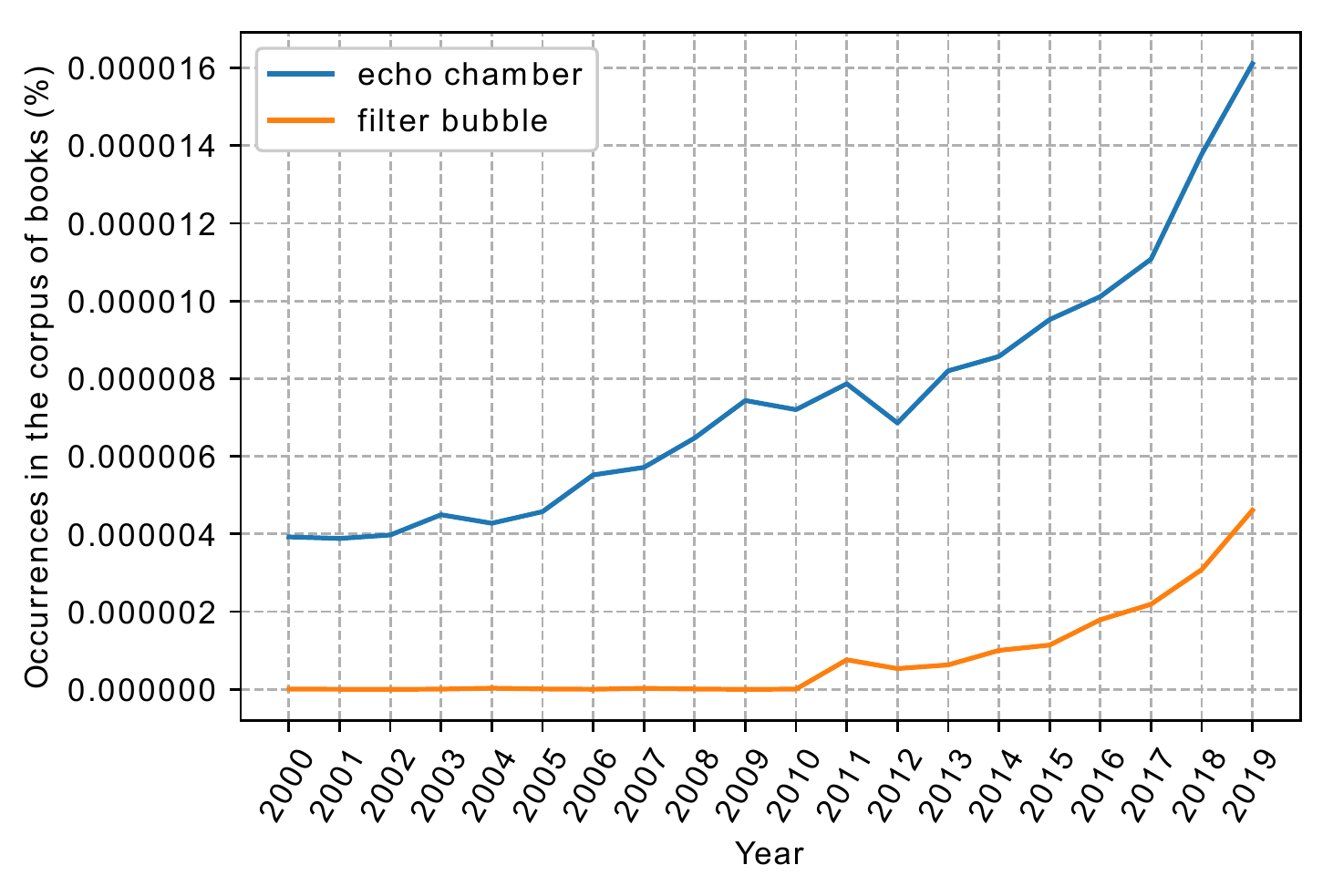}
\caption{Evolution of occurrences of the terms ``echo chamber'' and ``filter bubble'' from 2000 to 2019 in the corpus of books stored by Google in its digital database. 
The y-axis shows, of all the bigrams contained in Google's sample of books written predominantly in English and published in any country, what percentage of them are ``echo chamber" or ``filter bubble" (source: Google Books Ngram Viewer~\cite{misc_ngram_viewer}).}
\label{fig:ngrams}
\end{figure}

The number of mass media articles, scientific papers, and books about topics such as the increasing polarization and the strengthening of echo chambers and filter bubbles in social networks is growing year by year, as illustrated by Figure~\ref{fig:ngrams}. 
There have been hundreds of publications in the last two decades in this area. 

The goal of this annotated review is twofold. First, we identify the most used network polarization measures and the strategies used to handle the polarization problem, together with their main applications. 
Second, we present a comprehensive and annotated list of publications related to the evaluation of the polarization strength and to strategies used to handle the polarization.

The review is organized as follows. Section~\ref{sec:methodology} describes the methodology applied to collect the publications that propose or use network polarization measures or strategies to handle the polarization. It also includes the queries and digital libraries we used, together with some quantitative results. 
The polarization measures found in the reviewed papers are presented in Section~\ref{sec:measures}. 
Different approaches used to handle the polarization problem by using interventions, modifying the recommendation algorithms, or redesigning the network are described in Section~\ref{sec:reduction}. 
Section~\ref{sec:discussion} presents the concluding remarks, which also include short comments about real-life case studies and practical applications of polarization measures and models.

\section{Methodology}
\label{sec:methodology}

As a first step for creating this review, we conducted a systematic literature mapping~\cite{2012Wohlin}. We identified publications that propose or use network polarization measures or strategies to handle the polarization. 
This section presents the systematic mapping planning and the quantitative results we achieved. 

The universe of relevant analyzed publications consisted of journal and conference papers about the mathematical or computational modeling of network polarization. From the broad specter of publications dealing with polarization, we are specifically interested in those that present or use polarization measures, as well as in models that use them to handle polarization, from a theoretical or a practical perspective. 

Polarization may be understood as an existing phenomenon that can be modeled and measured. 
However, this concept can also be used to refer to the process of increasing the division between separate groups of individuals, parties, or other entities. In the first case, polarization can be seen as the state of some network. 
In the second, the polarization process is dynamic and subject to external modifications. These modifications, commonly known as external interventions~\cite{2018Gillani}, aim to bring the network to a new, different state. In most cases, the goal of an intervention is to reduce a specific polarization measure.
In this annotated review, we are also interested in publications that model these interventions, reducing or increasing some explicit or implicit polarization measure. 
Therefore, our two research questions were: 
\begin{enumerate}[label=Q\arabic*.]
    \item What approaches have been proposed for measuring network polarization?
    \item What network polarization reduction methods have been suggested?
\end{enumerate}


We chose the following set of initial keywords: polarization, network, measure, intervention, reduce, increase.

\subsection{Search string and digital library}



The Scopus digital library was chosen for retrieving the publications for this review. 
It is one of the most used digital libraries and provides access to journals, conference proceedings, and book chapters from ACM, IEEE, Springer, Elsevier, and other publishers. 
Compared to other digital libraries and search engines such as the Web of Science and Google Scholar, Scopus offers a good balance between a broad coverage of publication venues and their quality. Scopus indexes nearly the entire ScienceDirect database~\cite{2018ELsevier}. 
Web of Science is more restrictive when choosing the scientific journals it covers. 
On the other hand, Google Scholar includes many non-peer-reviewed sources. 

The Scopus advanced document search engine allows performing complex search queries using Boolean operators, approximate phrases, and field codes to narrow the scope of the search. Our search string was built using the initial keywords and using the above features, as presented below:

\begin{quotation}
TITLE-ABS-KEY ( \newline
    (polarization OR ``echo chamber*'' OR ``filter bubble*'') \newline
    AND (graph OR graphs OR network OR networks) \newline
    AND (metric* OR measure OR reduce* OR reduction OR increase OR intervention*)) \newline

AND ALL ( \newline
    (polarization OR ``echo chamber*''  OR ``filter bubble*'') \newline
    AND NOT ``cell network*'' 
    AND NOT antenna* 
    AND NOT radar 
    AND NOT ``cell polarization'' 
    AND NOT electromagnetic* 
    AND NOT electric 
    AND NOT optical) \newline

AND ( \newline
    LIMIT-TO(SUBJAREA, ``MATH'') \newline 
    OR LIMIT-TO(SUBJAREA, ``COMP'') \newline
    OR LIMIT-TO(SUBJAREA, ``SOCI'') \newline
    OR LIMIT-TO(SUBJAREA, ``MULT'')) \newline
    AND (EXCLUDE(SUBJAREA, ``CHEM'')) \newline
    AND (EXCLUDE(SUBJAREA, ``EART''))
\end{quotation}

In the first component of the search string, the TITLE-ABS-KEY field code is used for finding the most relevant search terms in the title, abstract, or keywords of the retrieved publications. Polarization, echo chamber, and filter bubble are the most used terms when describing the polarization phenomenon since they appear in almost every relevant publication in the field.
The analysis and the research questions presented above guided the selection of the rest of the keywords.

Several off-topic terms are enumerated in the second component of the search string. These terms represent different fields of research that generated a large volume of noise in our initial searches. The AND NOT boolean operator is used for removing the uses of the term polarization that are not related to our research questions, such as electromagnetic, electric, or optical polarization. These are terms commonly used in their specific research fields. We also excluded cell, antenna, and radar network mentions. The ALL field code also removes search results when other search fields such as journal title, conference name, or publisher name contain the excluded terms.

In the third and last component of the search string, the LIMIT-TO statement is used to narrow the scope of our search to venues containing Computer Science, Mathematics, Social Science, or multidisciplinary (at least one of these) among its subject areas. Using the EXCLUDE statement, we also excluded chemical and earth-science subject areas to refine the search results and further exclude noise.

The search string execution returned 405 publications at January 1st, 2022. We did not used the publication date as a filtering criterion. However, all these publications appeared between 1986 and 2021.

\subsection{Selection strategy}

Despite the refinement performed by the search string, it still returned many off-topic publications not relevant to this research. It was necessary to perform a selection process among the retrieved publications. To this end, a set of inclusion and exclusion criteria was established. 

The inclusion criteria used to select the publications chosen for appearance in the review among the 405 originally retrieved were: 
(1) the publication venue must be peer-reviewed;
(2) the publication 
must meet
the research questions; and
(3) the publication should involve a polarization measure or model.
The exclusion criteria were:
(1) white papers, theses, or technical reports were discarded;
(2) duplicated studies, i.e., work that appeared more than once with the same or similar titles and content (e.g., extended abstracts and full papers); and
(3) short abstracts of conference papers, without the full content. 

To ensure that the publications indeed met the research questions, we also excluded those that classify posts or texts as polarized or non-polarized, but do not make use of an underlying graph or network structure. 

The selection process consisted of two filtering stages. In the first stage, we only considered the title and abstract of each publication. The 405 publications retrieved using the search string were reduced to 91 after applying this first stage of the inclusion and exclusion criteria. 

In the second stage, we performed a full reading of all publications filtered by the previous stage. Each publication was reviewed by a second, different author in the second stage, who classified the publication and wrote the annotation. This second and final stage generated a set of 72 publications. The inclusion and exclusion criteria cited above were applied in both filtering stages. 

\begin{figure}[ht]
	\centering
	\includegraphics[width=0.64\textwidth]{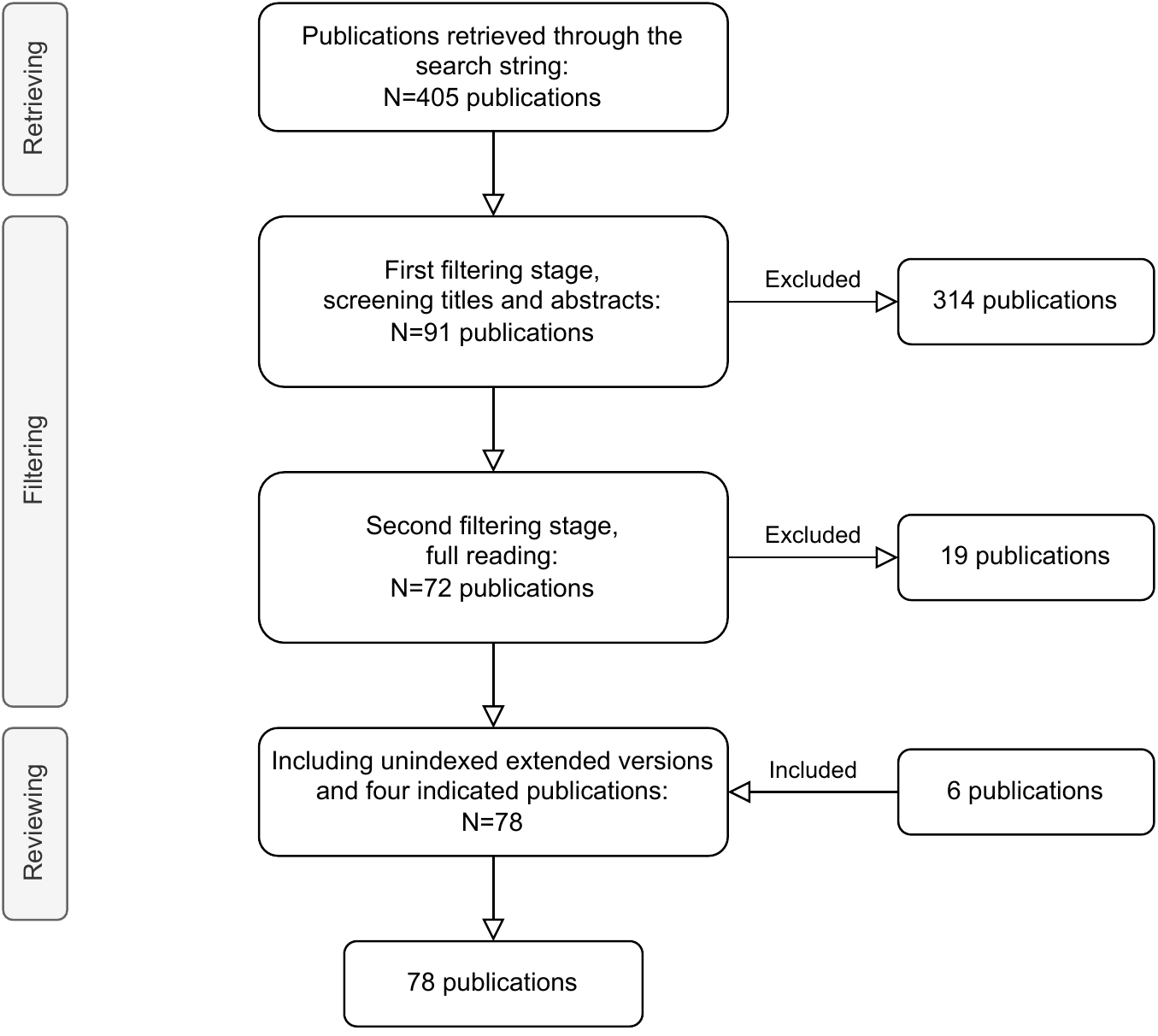}
	\caption{Flow diagram of the process of retrieving, filtering, and reviewing the publications.} 
	\label{fig:flow}
\end{figure}

We also added two additional unindexed (by Scopus) extended versions of publications that passed the filters. Lastly, we included 
four specific publications from 2022 indicated by the referees of this review. This process generated the 78 publications that are included in this review. Figure~\ref{fig:flow} shows the flow diagram of our methodology.

\subsection{Categorization of publications}

We created two main categories for classifying the publications according to the answers to the research questions: polarization measures and polarization reduction methods. An additional category of publications is formed by case studies and real-life applications involving the use of polarization measures and polarization reduction methods. 

\begin{figure}[ht]
	\centering
	\includegraphics[width=1\textwidth]{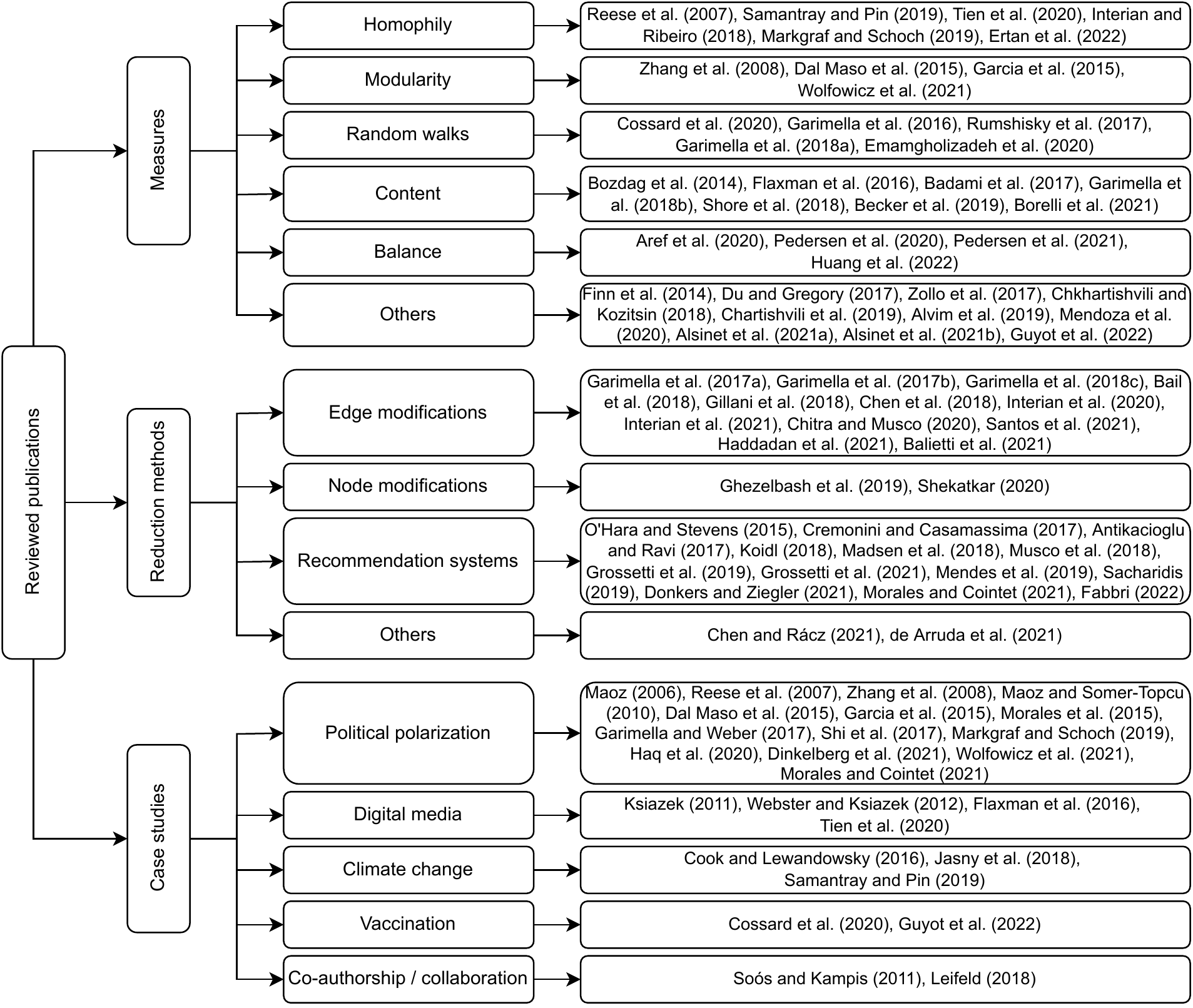}
	\caption{Categorization of the 78 reviewed publications.} 
	\label{fig:categ}
\end{figure}

One or more categories were assigned to each publication. 
Figure~\ref{fig:categ} shows the result of the categorization of the 78 reviewed publications. 

\subsection{Quantitative analysis}

This section provides a brief quantitative analysis of the 78 publications selected at the end of the two filtering phases, which will be simply referred to as ``publications'' throughout the remaining of this work. According to the Scopus document classification by type, most publications are journal articles (58.1\%), while conference papers represent 40.5\%, and the remaining 1.4\% correspond to book chapters.

\begin{figure}[ht]
	\centering
	\includegraphics[width=0.6\textwidth]{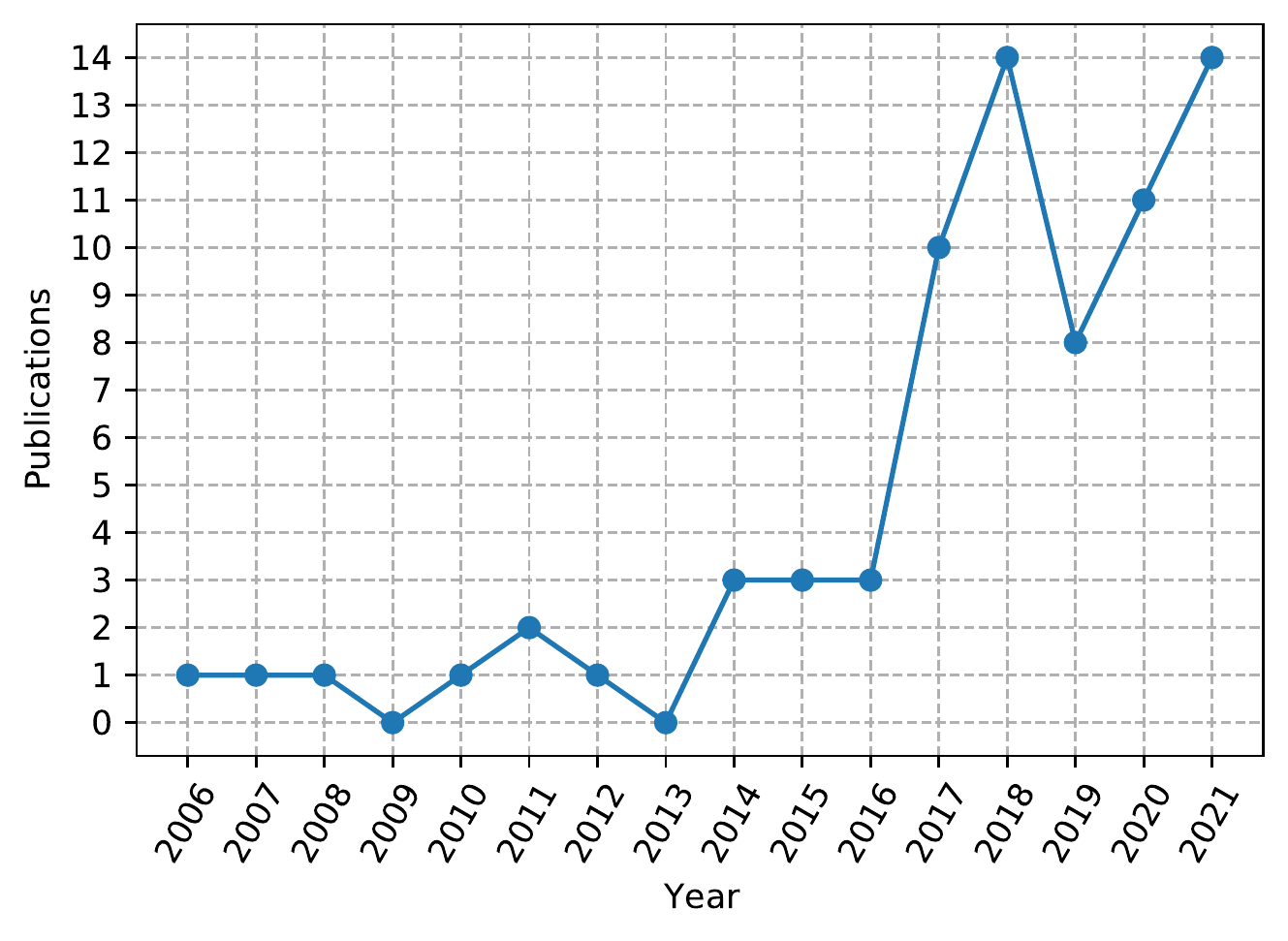}
	\caption{Publications by year (total of 73 publications, five publications of 2022 omitted).
}
	\label{fig:docs-by-year}
\end{figure}

Figure~\ref{fig:docs-by-year} details the number of publications by year since 2006. This number remains low and relatively stable until 2016, when it increases and reaches two peaks in 2018 and 2021, illustrating the increasing relevance and interest for the problem that occurs in parallel with the raising of the influence of debates in social networks such as Twitter, Facebook, and Reddit about society's issues. 

\begin{figure}[ht]
	\centering
	\includegraphics[width=0.84\textwidth]{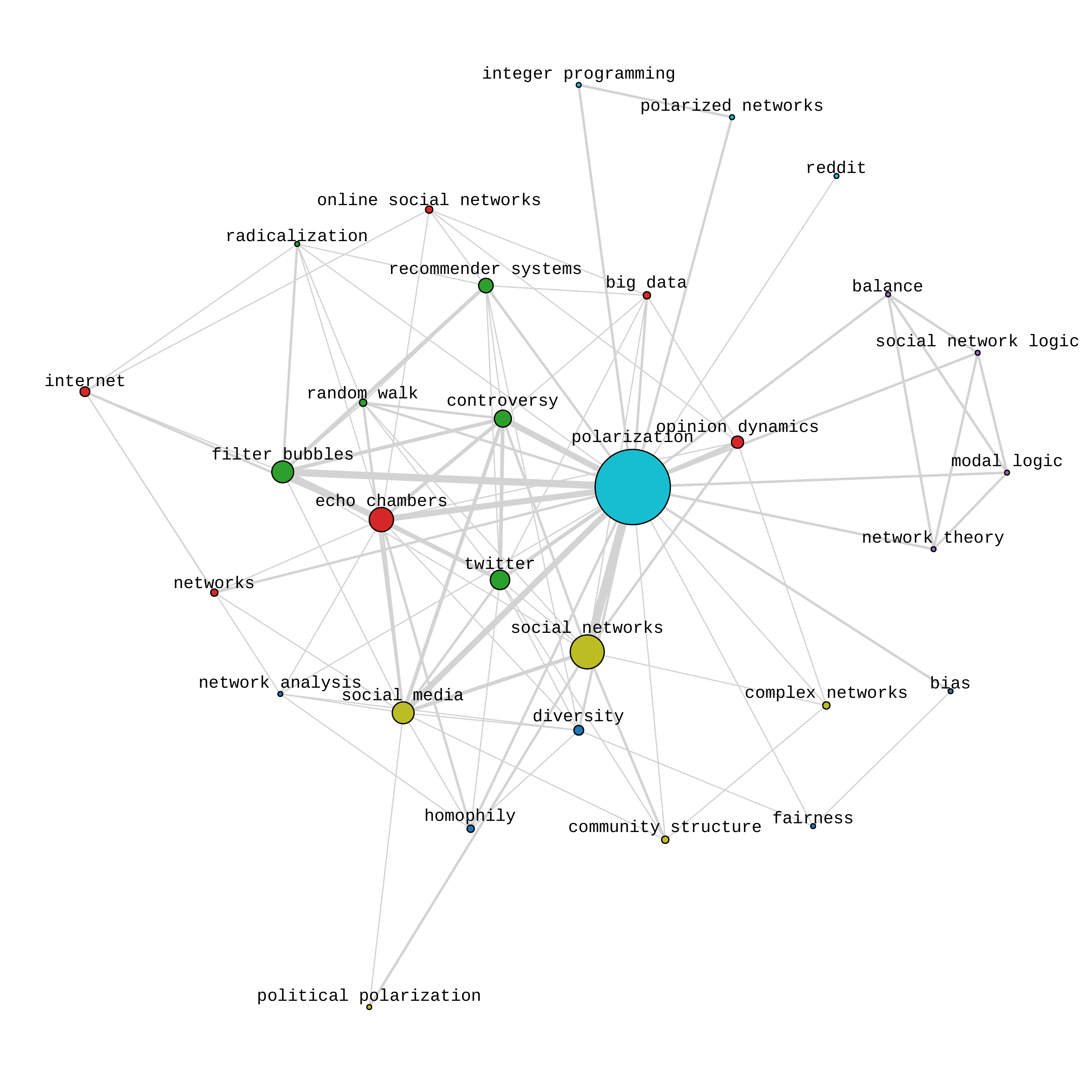}
	\caption{Co-occurrence network of the 30 author keywords with at least two appearances (78 publications).}
	\label{fig:co-occurrence-author-keywords}
\end{figure}

Looking at the 30 author keywords with at least two appearances in the reviewed publications, more than a half ($53.3\%$) of these 
appearances are concentrated in 
six terms: polarization (20.4\%), social networks (9.2\%), echo chambers (6.6\%), social media (5.9\%), filter bubbles  (5.9\%), and Twitter (5.3\%). 
The co-occurrence network of these 30 author keywords 
is shown in Figure~\ref{fig:co-occurrence-author-keywords}. Node sizes indicate the number of appearances of each keyword. ``Polarization'' is the most common keyword, with the highest centrality in the co-occurrence network and connected to virtually all other keywords. The keywords ``echo chambers'' and ``social networks'' are also frequently found. 

\section{Polarization measures}
\label{sec:measures}

This section presents the main approaches for measuring network polarization found in the reviewed publications.

The core of any polarization model lies in the concept of a group. Most of the time, group refers to any subset of network nodes that share some common characteristics. In some cases, groups are called communities. A strongly cohesive group that is loosely connected to other groups is commonly identified by the term echo chamber.

Let $G = (V, E)$ be a graph or a network, where $V = \{v_1,\ldots, v_n\}$ is its node set and $E 
\subseteq V \times V$ is its edge set. 
Let $A = \{A_1, \ldots, A_q\}$ be a set of node groups defined over $V$, i.e., each $A_i \subseteq V$ for any $i = 1, \ldots, q$. 

The group membership of some node $v \in V$ can be modeled in several ways. 
In the typical and more frequent case, groups form a partition of the node set, i.e., each node belongs to exactly one group~\cite{2016Garimella_Quantifying}. 
This case also allows the representation of nodes that are isolated or do not belong to any specific group. In these situations, we denote by $s(v)$ the only group to which node $v$ belongs. 
However, other situations might exist. For example, 
groups could form a cover of the node set, i.e., some nodes could belong to more than one group~\cite{2018Interian}.
Groups could also be modeled by fuzzy sets~\cite{1965Zadeh}, i.e., with the node membership to a group $A_i$ being given by a function $f_{A_i} \in [0,1]$. 
The higher the value of $f_{A_i}(v)$, the higher the grade of membership of node $v$ to $A_i$~\cite{2016Flaxman}.

Perhaps the most interesting and less commonly used membership model is that of fuzzy membership. An example of fuzzy membership can be found in audience fragmentation models: each consumer can divide its attention into several, even ideologically diverse, media outlets. Other examples of models that use fuzzy membership are content qualification methods (Section~\ref{sec:measure_content}), which use the content published or consumed by the users to measure their political leaning.

Understanding the polarization measures is crucial for applying the different approaches that model polarization. Understanding the measures is also important to assess the polarization reduction strategies that will be presented later in this review: any polarization reduction method must clearly establish the output that an intervention will reduce. 

Several approaches are described in the literature for measuring the polarization of a network. Although some of these approaches seem to be more consolidated and have been more frequently used in case studies and applications, this is an emerging field and, naturally, different approaches are considered. 
Therefore, Sections~\ref{subsec:homophily} to~\ref{sec:signed} present the most consolidated and used approaches in the literature, while other, more sparsely or less used approaches proposed for polarization evaluation are commented on in Section~\ref{other}.

\subsection{Homophily}
\label{subsec:homophily}

\textit{Homophily} (from Ancient Greek: \textit{homo} = ``self'' and \textit{philia} = ``love'', love to oneself) is the tendency of individuals to associate with others that are similar to themselves.
The strength of the homophily is directly related to the strength of the network polarization. 
An important advantage of assessing the homophily is that it can be measured at the node, group, or network level.
The term \textit{assortativity}~\cite{2003Newman} is often used for defining a similar concept to that of homophily. The difference between them is that, in practice, assortativity is often used for measuring the preference for similar nodes in terms of their degrees. 

The first use of the homophily for measuring the polarization of groups was made by Currarini et al.~\cite{2009Currarini} in the context of economic research. They used the term \textit{segregation} to refer to the separation or isolation by some criterion, a very close concept to that of polarization. Later, this concept was also extended to evaluate the polarization of specific nodes by Interian and Ribeiro~\cite{2018Interian}. 

Let the $i$-degree of a node $v$ of a network be the number $d_i(v)$ of neighbors of $v$ that belong to the group $A_i$. 
The homophily~\cite{2018Interian} of a node $v$ with respect to the group $A_i$ is defined as the ratio between its $i$-degree and the total number $d(v)$ of neighbors of node $v$:

$$
h_i(v) = \frac{d_i(v)}{d(v)}, \mbox{ for any } v \in A_i.
$$

This definition only makes sense if $d(v) > 0$, and the homophily is only defined for nodes that fulfill this condition, i.e., non-isolated nodes. The value of the homophily is a real number in the $[0,1]$ interval, where 0 suggests \textit{heterophily} (preference for the opposite), while 1 indicates extreme homophily.

The group-level homophily measure $H_i$ defined by Currarini et al.~\cite{2009Currarini} denotes the average $i$-degree of all nodes in the group $A_i$, divided by their average degree:



$$
H_i = \frac{\sum_{v \in A_i} d_i(v)}{\sum_{v \in A_i} d(v)}.
$$

The network-level homophily $H$ is similarly defined, considering all nodes in the network instead of those in a single group: 

$$
H = \frac{\sum_{v \in V} d_{s(v)}(v)}{\sum_{v \in V} d(v)}.
$$

The homophily is a simple statistical measure that indicates the strength of the ``preference for the similar'', a tendency present in many real-world networks. However, its simplicity comes with a set of drawbacks. For example, if the two groups $A$ and $B$ of a network divide nodes in the proportion of 90\%:10\%, then a group-level homophily measure of 0.5 can have a very different meaning for each group. For the small group, it seems to be very large. Contrarily, for the group representing the majority of the nodes, it appears to be insufficient to affirm that there is polarization. There are more refined polarization measures. 

Reese et al.~\cite{2007Reese} analyzed 
how polarized the major political blogs in the USA were, in terms of their homophily. They used a slight modification of the homophily measure, considering as neighbors nodes that are at a distance of one or two edges. 
Samantray and Pin~\cite{2019Samantray} studied a model of polarization of beliefs 
considering, in addition to the homophily, the degree of information credibility. 
They used a definition of group-level homophily~\cite{2009Currarini} and a polarization measure~\cite{2016Lelkes} that took into account the expressed opinion and the emotional content. 
Tien et al.~\cite{2020Tien} used 
principal component analysis~\cite{1901Pearson} to compute a left/right media score for each node. In this study, the assortativity measured the extent to which retweets occur between nodes with similar media preferences.

The publications below construct or use 
other polarization measures based on the definition of homophily. 

Interian and Ribeiro~\cite{2018Interian} 
%
analyzed the distribution of the homophily values over the nodes 
of a network
as an indicator of the strength of polarization, using a probabilistic approach to define a new homophily-based polarization measure. 
It consists on the calculation, for each node, of the probability 
of observing a number of same-type successors that is greater than or equal to the actual number of same-type successors observed for this node. This 
measure was used to assess the statistical relevance of the homophily value.
The authors also developed a probabilistic approach to compare the polarization of groups of nodes or entire networks,
based on the computation of empirical cumulative distribution functions of the sampled data from the network.
These empirical cumulative distributions provide a more insightful understanding of the status of the network. They may be used not only to compare the polarization of different groups of nodes or entire networks, but also to estimate the impacts of external interventions in terms of 
polarization.

Markgraf and Schoch
\cite{2019Markgraf} 
presented a 
framework for 
echo chamber research in online social networks. 
The first step 
lies in 
data collection. The second 
is community detection. The third is to assess the ideological views of the users. The fourth 
is to measure the degree of \textit{echofication}, i.e., to what degree a community qualifies as an echo chamber. 
A variant of the group-level homophily measure for multi-party networks was proposed,  
based on the cosine similarity between the users. 
The group-level homophily is calculated as the average similarity of the users to its neighbors in the network. 
A case study based on data from the 2017 German federal election to evaluate the framework is used. 
The authors argued that many researchers have reported the average homophily of the network’s users, but neglected separate groups that host particularly polarized users. 
According to the them, this approach has led to underestimating the polarization of communities that host groups of political extremes.

Ertan et al.~\cite{2022ERTAN}
argued that there is a well-established literature on measuring political polarization in two-party systems, but very limited for multi-party systems.
They proposed 
measuring polarization for multi-party systems from survey studies. 
A cognitive political network (CPN) is generated for each respondent of the survey by asking them how they perceive the relationships between each possible pair among all $n$ major political parties. 
From the CPN, measures of 
perceived party polarization were calculated. 
The $\mbox{E-I}=\frac{ET - IT}{ET + IT}$ index~\cite{1988Krackhardt} 
proposed in the context of social psychology was used, where $ET$ is the number of inter-group edges and $IT$ is that of intra-group edges.  
It ranges between -1 and 1: values close to -1 indicate the network is dominated by intra-group edges (homophily), and values close to 1 show the presence of heterophily, i.e., extensiveness of inter-group edges.
This measure corresponds to one minus twice the value of the homophily.

\subsection{Modularity}
\label{sec:modularity}

Modularity optimization is a well-known method for community detection~\cite{2006NewmanA}. 
It treats community detection as an optimization problem by seeking an assignment of nodes to communities 
that maximizes 
some objective function. However, the modularity has also been used for measuring the polarization inside the groups.

The \textit{modularity} evaluates the number of intra-community against inter-community edges for a given set of node groups in a network. 
In short, the modularity is, up to a multiplicative constant, 
the number of intra-group edges in the network minus the expected number of intra-group edges in a network with the same nodes, communities, and node degrees, but with edges placed at random.
The mathematical definition of the modularity 
was originally proposed by Newman~\cite{2006NewmanA}: 
$$
Q = \frac{1}{2|E|} \sum_{u,v \in V}(a_{uv} - \frac{d(u) d(v)}{2|E|}) \cdot g(u,v),
$$
\noindent
where $|E|$ is the 
number of edges in the network; 
$d(v)$ is the degree of node $v \in V$;
$a_{uv} = 1$ if there is an edge between nodes $u$ and $v$, $0$ otherwise; 
and $g(u,v) = 1$ if $u$ and $v$ belong to the same group, $0$ otherwise. 

Some publications used the modularity in case studies that investigated political polarization. 
Zhang et al.~\cite{2008Zhang} used the modularity to quantify the increase of polarization in the US congress in the period 1979-2004. They identified communities of congressmen by employing a slight modification of the leading-eigenvector community detection method~\cite{2006NewmanB}. 
Dal Maso et al.~\cite{2015DalMaso} 
used the modularity to 
evaluate the 
polarization between government and opposition in the Italian parliament.
It was defined as the average modularity decrease after the group swap of two opposite-group nodes, calculated over all pairs of opposite-group nodes. The larger the decrease in the modularity, the larger is the polarization between the two groups. 
Garcia et al.~\cite{2015Garcia} presented an empirical analysis of 
politnetz.ch, a Swiss online platform focused on political activity. 
The approach 
focused on
the construction of a multiplex network with politicians as nodes and three layers of directed links: one with support links, a second one with link weights as the amount of comments a politician made to another politician, and a third one with weights counting the amount of times a politician liked the posts of another. 
The polarization was studied in the three layers 
and measured as the modularity with respect to party labels. 
Wolfowicz et al.~\cite{2021Wolfowicz} used the modularity as one of the polarization measures in a study about the interactive effects of filter bubbles and echo chambers on radicalization. 


\subsection{Random walk controversy}

The \textit{random walk controversy} (RWC)~\cite{2016Garimella_Quantifying,2018Garimella}
is defined as follows. Given 
network $G=(V,E)$, let $X,Y \subseteq V$, with $V = X \cup Y$, define a partition of its node set into two subsets. 
Consider two random walks, one ending in $X$ and the other in $Y$. The random walk controversy measures the difference between two probabilities: 
\begin{itemize}
  \item $Pr[A]$: probability that both random walks started from the same partition where they ended.
  \item $Pr[B]$: probability that both random walks started from different partitions than 
  they ended in.
\end{itemize}
Then, the
$$RWC = Pr[A] - Pr[B]$$

\noindent
is close to one when the probability of crossing sides is low, i.e., when the graph is polarized. On the other hand, it is close to zero when the probability of crossing sides is comparable to that of staying on the same side, meaning that there is no polarization. 

There is a variant of the RWC measure~\cite{2018Garimella} that can be computed more efficiently. It only considers random walks that end when reaching any of the $k$ highest-degree nodes from either partition. In this variant, the high degree is used as a proxy for authoritativeness. The sets of the $k$ highest-degree nodes of each group are denoted by $X^{+}$ and $Y^{+}$. 

The node-level RWC 
of a node $v \in V$ with respect to the group $X$ 
considers how often a random walk originating at $v$ ends in nodes from $X^{+}$ and $Y^{+}$. Formally, it is defined as the probability that a random walk started at node $v$ given that it ended in $X^{+}$, divided by the sum of the probabilities that a random walk started at node $v$ given that it ended in $X^{+}$ or in $Y^{+}$. The measure


$$
RWC(v, X) = \frac{Pr[start=v~|~end=X^{+}]}{Pr[start=v~|~end=X^{+}] + Pr[start=v~|~end=Y^{+}]}
$$

\noindent
is close to one if node $v$ is located near the highest-degree nodes $X^{+}$ in the network, while it is close to zero when $v$ is far from $X^{+}$ and close to $Y^{+}$.

Cossard et al.~\cite{2020Cossard} analyzed the Italian vaccination debate on Twitter, using the random walk controversy measure for quantifying the polarization. 

Garimella et al.~\cite{2016Garimella_Quantifying,2018Garimella}
presented and compared three new
and two existing polarization measures. 
They argued that the random-walk-based measure outperforms other measures in capturing the intuitive notion of controversy, which is the concept used in this work to refer to polarization. To build graphs from raw data, the authors used ``retweet'' and ``follow'' relations between Twitter users. 
A three-stage pipeline that leads to quantifying controversy in any network is also proposed:
build a conversation graph among the users who contribute to a topic, where an edge represent 
two users 
in agreement; partition the conversation graph to identify potential sides of the controversy; and measure the amount of controversy from characteristics of the graph. 
It is claimed that the RWC is able to separate controversial from non-controversial topics 
and
that this 
score can be used to generate recommendations that foster a healthier ``news diet'' on social media.

Rumshisky et al.~\cite{2017Ciampaglia}
looked at the 
RWC measure,
text-based sentiment analysis, and the corresponding shift in word meaning and utilization by the opposing sides, using the 2014 Ukraine-Russian Maidan crisis as a case study. 
They analyzed the interplay of the division of network-based vs. language-based measures of conflict, using the RWC as a network measure and the standard deviation of sentiment and semantic drift as verbal measures. 
They observed that, as the conflict intensifies, 
the RWC 
and the standard deviation of the overall sentiment expressed by the opposing groups are positively correlated and  increase in unison.)

Emamgholizadeh et al.~\cite{2020Emamgholizadeh}
sought to determine to what extent an idea produced by 
some
user is exposed to members with an opposite point of view. 
They introduced the Biased Random Walk (BRW), a method for combining two sources of information: content or textual data and structural data. 
Content and structural network data are used by the biased random walker, which has some amount of initial energy (calculated considering the position of the node from where the walk starts) that is loosed in each step along the walk. The performance of BRW is compared with that of a pure random walk~\cite{2016Garimella_Quantifying,2018Garimella}. They argued that, in some cases, using only structural data, it is not possible to evaluate the controversy level of the social network. 

\subsection{Content qualification}
\label{sec:measure_content}

Content qualification methods use the content published or consumed by the users to measure their polarity. 
They do not consist of a single method: instead, each publication defines in its own way how the features of the published or consumed content will be transformed into the polarity of the nodes of the network. 
However, all of them employ user content (hashtags, web links, sentiment analysis) for identifying the group membership or the polarity of network nodes. 

Bozdag et al.~\cite{2014Bozdag} 
analyzed pluralism 
on Twitter, measuring information diversity in 
Netherlands and Turkey. 
They coined the term \textit{cyberbalkanization} to refer to the internet segregation into small groups with similar interests (i.e., polarization). 
Several metrics were used, among which we highlight three: source diversity, output diversity, and input-output correlation. 
Entropy is used to assess the diversity of information consumed or produced by each user. 
Source diversity
is measured 
by the entropy of the tweets published by followed users from different groups. 
Output diversity is estimated from the entropy of the retweets and replies the user makes. 
The 
input-output correlation indicates whether the political position of the most common message category retweeted by a user is significantly skewed from the political position of the received messages. 
The results indicated a high source diversity, similar for Turkish and Dutch users. 
The output diversity is much lower than the source diversity. 
Considering 
the minority access, 
the content produced by minorities cannot reach a large fraction of the Turkish population.

Flaxman et al.~\cite{2016Flaxman}
examined the web browsing patterns of 50,000 
US-located internet users who regularly read online news, 
defining four channels individuals use to discover a news story: direct, aggregator (Google News), social (Facebook, Twitter), and search (web search queries on Google, Bing, Yahoo). 
They 
estimated the polarity of a news outlet by measuring how its popularity varies across counties as a function of their political compositions. 
A Bayesian model was used for estimating the polarity of each article and user. 
The polarity score of each article is inferred from the polarity of its publisher.
Using the polarity scores of the articles read by some user, the model also estimates their polarity. 
The segregation (i.e., the ideological distance) between two individuals is defined as the expected value of the squared distance between their 
polarity scores. 
It was shown that articles found via social media or web search engines have higher ideological segregation than those users read by directly visiting news sites. 
However, it was also found that social media and web search engines are associated with greater exposure to opposing perspectives.

Badami et al.~\cite{2017Badami}
observed the importance of understanding how recommendation systems behave in polarized environments,
studying polarization in the context of the users' interactions with a space of items. 
Their model works with ratings that 
capture the distribution of user opinions. 
In the absence of polarization, the distribution of opinions should be either J-shaped or bell shaped. 
As polarization emerges, the 
distribution becomes U-shaped, with two peaks emerging around the two 
confronted opinions at the extreme sides of the rating scale. 
The authors developed an approach to quantify polarization based on four stages: (1) building items' rating histograms from user-item rating data; (2) extracting a set of features from the histograms; (3) training a polarization classifier based on a sample of annotated cases; and (iv) measuring the item-level polarization score. 
They 
performed comparisons of 
polarization measures on several 
benchmark datasets 
and showed that their framework can detect different degrees of polarization.

Garimella et al.~\cite{2018Garimella_wwwconf} 
assessed the degree to which echo chambers are present in political discourse on Twitter and how they are structured in terms of 
different user roles. 
Two node-level measures related to how polarized is the content that each user consumes and produces are defined. 
The production polarity of an user is 
the average polarity of its tweets. 
The consumption polarity of an user is 
the average polarity of the tweets it receives from the followed users. 
A user is classified as partisan if it produces one-sided content, and bipartisan if it produces two-sided content. 
The authors also looked at gatekeeper users, who consume content of both leanings, but produce 
single-sided
content. 
The findings indicated a large correlation between the leaning of produced and consumed content. 
Partisan users enjoy a higher ``appreciation'' as measured by network and content features, indicating a ``price of bipartisanship'', paid by users who try to bridge echo chambers. 
They pay the price in terms of network centrality and endorsements from other users, highlighting the existence of a latent phenomenon that effectively stifles mediation between the two sides. 

Shore et al.~\cite{2018Shore} 
sought evidence of echo chambers strengthening by analyzing the diversity of hyperlinks posted on Twitter. 
They used ordinary least squares models to test their hypotheses, combined with standard community detection algorithms to identify the groups. 
They found that the average account posts links to more politically moderate news sources than the ones they receive in their own feed. 
However, members of a tiny network core do exhibit cross-sectional evidence of polarization and are responsible for the majority of tweets received overall due to their popularity and activity, which could explain the widespread perception of polarization on social media.
Evidence was also found that people in highly clustered positions (echo chambers) tweet more similarly to the people they follow.

Becker et al.~\cite{2019Becker} 
analyzed  
partisan networks of Republicans and Democrats to test whether the wisdom of crowds is robust to partisan bias. 
They studied belief formation on controversial topics. 
Two web-based experiments were conducted, where each individual answered 
questions to elicit partisan bias before and after observing the estimates of peers (social information) in a politically homogeneous social network. 
Polarization was measured using two outcomes. 
First, the average distance (absolute value of the arithmetic difference) between the mean normalized belief for Republicans and Democrats at each experiment round. 
Second, the average distance between every possible two-person cross-party pairing, which reflects the expected distance between the beliefs of 
randomly selected Democrats and 
Republicans. 
The experimental results indicated that social information in politically homogeneous networks do not always amplify existing biases. 
Instead, in the studied networks, the information exchange increased belief accuracy and reduced polarization.

Borelli et al.~\cite{2021Borrelli} 
%
examined the relationship between online emotional reactions, affective polarization, and counter narratives, following an approach based on user-generated textual data. 
Affective polarization is the extent to which two opposing groups dislike one another. 
This occurs in online social networks as a result of a controversial event in the offline world. 
A content-based measure of affective polarization derived from user-generated content was proposed 
to evaluate the usage of a set of controversial words, 
reflecting how important a word is to a document in a collection. 
Also,
a new method is proposed to assess the effectiveness of online counter narratives made by influential actors to counteract the rise of affective polarization. 
It was applied to five cases of controversial events that occurred in European soccer leagues using 
Twitter data, 
showing that there was a high polarization in online responses in most cases. 
Counteractive official communication from the clubs within 12 hours of the event often reduced the affective polarization.

\subsection{Signed networks and balance theory}
\label{sec:signed}

The notion of balance comes from the idea that, in a group of people, some logical rules are generally observed (e.g., people like their friends' friends, people hate their friends' enemies). If a social network always satisfies these rules, it is said to be balanced. 

Cartwright and Harary~\cite{1956Cartwright} and Heider~\cite{1946Heider} studied the theory behind such 
relationships and attitudes. 
The network is modeled as a signed graph $G(V,E^+,E^-)$, which consists of a set $V$ of vertices and two disjoint subsets $E^+,E^-$ of positive and negative edges, respectively. 
Formally, balance is achieved whenever each triangle (or 3-cycle) has three positive edges (\textit{my friend's friend is my friend}) or two negative and one positive edge (\textit{my friend's enemy is my enemy}). 

The structure theorem~\cite{1956Cartwright} shows that a signed graph is balanced if and only if its nodes can be separated into two disjoint subsets such that each positive edge joins two nodes of the same subset, while each negative edge joins nodes from different subsets. 
Balanced graphs may be used as a model of polarized networks. A graph with exactly two groups of nodes linked internally only by positive edges and with negative edges between the groups represents a perfectly polarized network. 
In the case of weak balance, 
the existence of triangles with three negative edges is also allowed. 
Weak balance~\cite{2020Pedersen,2021Pedersen} is characterized by the possibility of partitioning the nodes into any number of groups. 

Harary~\cite{1959Harary} defined some measures for evaluating how close a given graph is to balance. 
The degree of balance of a signed graph $G$ 
is given by the ratio of the number of positive cycles to the total number of cycles, where the sign of a cycle is the product of the signs of its edges: 
$$ \beta(G) = \frac{c^{+}(G)}{c(G)}.$$

The line index of balance is the minimum number of edge modifications that must be made in order to achieve balance. The two most used 
modifications are edge removals and sign changes. 
The third measure is the point index, given by the smallest number of nodes whose deletion results in balance.

Aref et al.~\cite{2020Aref} 
investigated the relationship between network structural configurations and tension in social systems by using balance theory
and three levels of analysis for balance assessment: triads, groups, and the whole network, delivering empirical evidence for the argument that balance at different levels represents different network properties
that should be evaluated independently. 
For triad-level balance, a new measure was developed by using semicycles that satisfy the condition of transitivity. 
For group-level balance, the measures of cohesiveness (intra-group solidarity) and divisiveness (inter-group antagonism) were proposed to capture balance within and among groups. 
For network-level balance measurement, the authors modified the line index of balance~\cite{1959Harary}, 
introducing a normalized line index. 
Large values of this index represent high partial balance, and therefore a more balanced network. 
The investigation of different social networks 
showed that balance appeared differently across multiple levels of analysis. 
In most cases, relatively high values of balance were observed, corresponding to high triad, group, and network polarization. 

Pedersen et al.~\cite{2020Pedersen,2021Pedersen}
proposed different ways of defining properties related to the concept of balance in signed social networks, where relations can be either positive or negative. 
To be able to formally reason about the social phenomenon of group polarization based on balance theory, 
they used positive and negative relations logic~\cite{2020Xiong}. 
Positive and negative relations between nodes are interpreted as agreement or disagreement on a given issue. 
They studied a polarized network as a balanced graph of groups positively related within, but negatively related to the others.
They differentiated strong and weak polarization. Strong polarization occurs when the network can be divided into two mutually opposed groups. Weak polarization is characterized by 
the division in many groups.
The authors presented three measures: degree of imbalance, level of imbalance, and line index of imbalance, 
defined for both the strong and weak polarization cases. 
They 
discussed their strengths and weaknesses 
on
examples of signed networks.

Huang et al.~\cite{Huang2022} focused on predicting conflicts as negative links between users. According to the authors, negative links between polarized communities are too sparse to be predicted by state-of-the-art approaches. A polarization measure for signed graphs that incorporates social balance theory is proposed based on signed random walks. 
This measure guarantees polarized similarity consistency, satisfying two properties: (1) topologically close nodes are more similar than topologically distant ones, and (2) positively related nodes are more similar than negatively related ones. 
Then, POLE (POLarized Embedding for signed networks), a signed embedding method for polarized graphs based on random-walk based measure 
is proposed. Through the experiments, the authors claimed that POLE outperforms state-of-the-art methods in hostile links prediction. 

Table~\ref{tab:measures} shows the polarization measures 
found in the reviewed publications and discussed in Sections~\ref{subsec:homophily} to~\ref{sec:signed} that can be used for assessing the polarity of individual nodes, or the polarization of groups or entire networks. 
Most measures aim to evaluate the polarization strength at the node or network level. 

\begin{table}[ht]
\centering
\begin{tabular}{lcccc}
\cline{1-5}
\multicolumn{1}{l}{Measure} & \multicolumn{1}{c}{\hspace*{0.1cm}node-level?} & triangle-level? & \multicolumn{1}{c}{\hspace*{0.1cm}group-level?} & \multicolumn{1}{c}{network-level?} \\ \hline
Homophily  & Yes & - & Yes & Yes \\
Modularity & -   & - & -   & Yes \\
Random walk controversy & Yes & - & - & Yes \\
Content qualification methods & Yes & - & - & - \\ 
Balance-based measures & - & Yes & Yes & Yes \\ \hline
\end{tabular}
\caption{Comparison of the
polarization measures in terms of their granularity.}
\label{tab:measures}
\end{table}

\subsection{Other approaches}
\label{other}

There are other, less used methods for polarization measuring that do not fit in the more frequently used approaches exposed in the previous sections. 
There are many different terminologies, methodologies, and measures, some of them similar to others. 
This section discusses other measures that use alternative techniques and ideas for quantifying node, group, or network polarization. 

Finn et al.~\cite{2014Finn}
%
introduced the co-retweeted network, 
which is the
weighted graph that connects highly visible accounts 
retweeted by members of the audience during some real-time event.
When applied to political conversations related to some event, the co-retweeted network enables the measurement of the political polarity of major players (including news outlets), based on the views of the audience on Twitter. 
Its first application is the measurement of the opinion polarization for an issue or topic, i.e., the computation of the polarity of the event itself. 
The second consists in measuring the media bias, as perceived by the Twitter users. 
The authors used their method to infer the polarity of all engaged 
accounts in the audience, in order to answer the question of whose 
supporters were more active and vocal during an event.

Du and Gregory~\cite{2017Du} 
%
investigated whether social media platforms increase the polarization of users, by checking if the community structure becomes stronger as time passes. 
Twitter networks that consider only reciprocated ``follow'' relations between users were used. 
The authors measured how often new edges appeared and whether edges tend to be removed (by ``unfollowing'') inside or between communities. 
Two hypotheses were explored:
(1) new edges are more likely to appear inside communities than between communities; and (2) edges between communities are more likely to be removed than those inside them. 
These two hypotheses were contrasted with the null hypothesis, when edges are added and deleted randomly. 
The authors showed that the number of intra-community edges added in the real networks is always much greater than in the random case. 
They also showed that inter-community edge deletion is more common than expected in the random case. 
Therefore, the polarization of the ``follow'' network becomes stronger. 
The authors argued that one possible explanation for this effect is the recommendation system of Twitter. 

Zollo et al.~\cite{2017Zollo}  
examined the effectiveness of debunking on Facebook through a quantitative analysis of 54 million users 
from January 2010 to December 2014. 
Debunking posts strive to contrast misinformation spreading by providing fact-checked information to specific topics. 
The authors compared how Americans 
who consume proven (scientific) and unsubstantiated (conspiracy-like) information on Facebook interact with 
debunking posts. 
Their findings confirmed the existence of echo chambers where users interact primarily 
with either conspiracy-like or scientific pages. 
The user polarity is defined as the ratio of the difference in likes (or comments) on conspiracy and scientific posts. The probability density function of the polarity of all users is sharply bimodal, and most users may be divided into these two groups. 
The majority of 
likes and comments is made by users polarized towards science, while only a small minority is made by users polarized towards conspiracy. 
Only few users active in the conspiracy echo chamber interact with debunking information.

Chkhartishvili and Kozitsin~\cite{2018Chkhartishvili} 
%
proposed the Binary Separation Index 
to quantify the echo chamber effect in 
social networks for a 
specific
topic. It requires the ideological space to be binary. 
It does not require the information of all users' opinions on 
the 
topic, but only of a subset of accounts that disseminate information in the network and their political positions. 
For a given 
social network and a fixed topic, it generates a number between 0 and 1: the higher it is, the greater is the level of information separation between the groups. 
However, the authors have not considered all possible information spreaders, not examining group pages. 
They 
discussed the calculation of this index for the prevalent Russian 
social network VKontakte. 
Considering the attitude towards the Russian government as a topic, they obtained 
an index
of 0.802 
after 
data processing, which was an evidence of a high level of information
separation among VKontakte users.

Chartishvili et al.~\cite{2019Chartishvili} 
proposed an extension of the Esteban-Ray measure~\cite{1994Esteban} originally proposed for measuring economic characteristics of a population. 
It may be applied when opinions are evaluated by continuous scalar values representing personal attitudes towards a fixed topic. 
The proposed extension evaluates the level of polarization of the individuals' opinions in a social network. 
An individual's opinion is described by a scalar in the interval $[0,1]$, representing the degree to which the individual holds a particular position.
The proposed polarization index is proportional to the difference between the average opinions inside the groups 
and belongs to the interval $[0,1]$. 
It is sensitive to the cluster sizes and reaches its maximum value when the groups are equal in size. 
The authors showed how the measure works for real data, applying it to a time series of user opinions in the VKontakte social network that are devoted to a political topic, reporting an increase in the level of polarization.

Alvim et al.~\cite{2019Alvim} 
%
also developed an Esteban-Ray-based polarization measure~\cite{1994Esteban} and a social network model. 
The model includes information about each agent’s quantitative strength of belief on a proposition 
and a representation of the strength of each agent’s influence on every other agent. 
The authors considered how the model changes over time as agents interact and communicate. 
They included several different options for belief update, such as rational belief update and update taking into account irrational responses such as confirmation bias (groups may strengthen their beliefs by interpreting information in their favor) and the backfire effect (groups may strengthen their beliefs by strongly opposing an opinion if it contradicts their views). 
The authors considered the evolution of polarization over time under various scenarios, as well as the implications of these results for real-world social networks.
Simulations were shown exploring how interaction graphs and cognitive biases may lead to polarization. 
Their experiments allowed to identify that 
sometimes people with different opinions interacting more strongly may lead to more polarization.

Mendoza et al.~\cite{2020Mendoza},
introduced GENE (Graph Generation conditioned on Named Entities), a representation of user networks conditioned on the entities (personalities, brands, organizations) that users comment upon. 
The goal was the early detection of polarization and controversy in news events. 
GENE segments the user network, and the segmented network is used to study two controversy indices, the Random Walk Controversy~\cite{2016Garimella_Quantifying,2018Garimella} and the Relative Closeness Controversy (RCC) proposed here. 
To evaluate the performance of GENE, the network of users of the online news site Emol~\cite{misc_emol} in Chile was modeled. 
The results showed that over 60\% of the user comments have a predictable polarity,
allowing both controversy indices to detect the controversy successfully. 
The authors argued that the RCC index shows satisfactory performance in the early detection of controversies using the information collected during the first hours of the news event. 
A polarization dynamic can be anticipated, predicting the emergence of controversies before they occur. 

Alsinet et al.~\cite{2021Alsinet_GNN,2021Alsinet} 
introduced a quantitative model for measuring polarization in online discussions. 
They modeled the debates in Reddit using weighted graphs with labeled edges, where node weights represent the polarity of the users' opinions in the debate, and edge labels represent the sentiments between users' opinions. 
The proposed measure is based on the maximum polarization of a debate, considering all possible graph bipartitions. 
For each bipartition, the polarization is quantified by measuring the uniformity of the users' opinions within each partition and the negativity of the interactions between the partitions. 
The maximum polarization is computed by a greedy local search algorithm. 
The authors argued that their approach can be used for monitoring a discussion and generating a warning signal when the polarization of the debate reaches some threshold value. 
They 
performed empirical evaluations of different Reddit discussions. 
The quantitative model captured differences in the polarization of different discussions. 
Additionally, a graph neural network~\cite{2017Hamilton} was used to approximately compute the polarization measure of a Reddit debate.

Guyot et al.~\cite{Guyot2022} stated that users in the boundaries significantly contribute to network polarization, acting like gatekeepers of their communities. They used an approach that relies on community boundaries to compute two measures: community antagonism and the porosity of boundaries. These measures assess the degree of opposition between communities and their aversion to external exposure, respectively. The authors evaluated their proposal using a case study obtained from Twitter and related to COVID-19 vaccination. 

\section{Polarization reduction}
\label{sec:reduction}

The second research question 
targeted network polarization reduction methods suggested in the literature, which are now exposed in this section.
All of them have in common some attempt of changing different features of the network: add or remove edges~\cite{2017Garimella,2021IntMorRib}; introduce specific types of nodes (zealots~\cite{2020Shekatkar}, informed agents~\cite{2019Ghezelbash}); or the spread of random information~\cite{2017Cremonini}. The publications may propose methods to compute optimal interventions or analyze the impact of such modifications in the network structure, evaluating their effect on the polarization of the entire network or on the polarity of specific nodes. 

Surprisingly, no studies about removing (or adding) nodes for reducing the polarization were found in this review. However, this method is often used in practice for banning specific posts or accounts from social networks~\cite{TheConversationBanning}. 

We observe that polarization reduction strategies seem to be more effective on users (nodes) that are new to the network~\cite{2018Madsen} or when a polarized discussion first emerges~\cite{2021Donkers}, i.e., when the polarization process is in formation, but not when it is already consolidated and the polarized groups are well established. 

\subsection{Edge modifications}

This approach, proposed in many publications, suggests that adding, removing, or changing the weight (or the strength) of some specific edges 
can reduce the polarization of the network. 
These 
interventions represent externally-induced processes that promote edge appearances or deletions, such as marketing or fact-checking campaigns, regulatory actions, or  direct manipulations that add or remove edges of the network. 

Most 
studies suggest that adding edges that link different groups may decrease group polarization. These edges usually represent the exposure of individuals to different or opposing views. However, Bail et al.~\cite{2018Bail} stated that this exposure to opposing views on social media may increase 
the polarization in some cases.

These are not necessarily conflicting hypotheses, since the exposure to opposing views may decrease polarization in the initial or intermediate phases of the process of  polarization, but not when the polarization is already strong. 
The effectiveness of this approach may depend on several factors. 
Among other researchers that tested edge additions in real-world networks, Cossard et al.~\cite{2020Cossard} argued that exposure to contrarian content has been shown to be both effective~\cite{2017Garimella,2015Horne} and counterproductive~\cite{2018Bail,2010Nyhan} in reducing the polarization, depending on the specific network setting or the existing degree of polarization. 


Garimella et al.~\cite{2017GarimellaA} 
elaborated a demo providing automated tools to help users explore and escape their echo chambers. 
A discussion topic is identified as the set of tweets that 
contain a specific hashtag. The topic is represented by an endorsement graph, where nodes represent users and edges represent endorsements. 
The two sides of a controversial topic are identified by employing a graph partitioning algorithm, dividing the graph into two subgraphs. 
Polarity scores are obtained for all users. 
The demo provides contrarian content recommendations, i.e., content that expresses views from the opposing side of the controversy. 
However, not all recommendations are acceptable, especially if they do not conform to the users' beliefs. 
To reduce these effects, an acceptance probability is defined, quantifying the degree to which a user is likely to endorse the recommended content. 
The maximum reduction of the user-polarity score is achieved by putting the user in contact with an authoritative source from the opposing side. 
The authors claim that the contribution of their demo is twofold. First, as a tool to visualize retweet networks about controversial issues on Twitter. Second, as a solution proposal to reduce the polarization by exposing users to 
contrarian views.

Garimella et al.~\cite{2017Garimella,2018Garimella_Reducing} 
studied algorithms 
for bridging the echo chambers created on social media, and thus reduce controversy (i.e., polarization). 
They represented the discussion on a controversial issue by an endorsement graph, raising an edge-recommendation problem on this graph. 
The goal of the recommendation is to reduce the random-walk controversy (RWC) score of the graph~\cite{2016Garimella_Quantifying,2018Garimella}. 
The authors also took into account the acceptance probability of the recommended edge. 
The goal is to find edges that produce the largest expected reduction in the controversy score. 
They 
proposed an algorithm that considers only 
edges between high-degree nodes of each side of the controversy. 
For each 
edge, 
it computes the reduction in the RWC score obtained when that edge is added to the original graph, 
then selects the $k$ edges that lead to the lowest scores when added to the graph individually. 
Experimental results showed that the 
algorithm is more time-efficient 
than a simple greedy heuristic, while producing comparable RWC score reduction. 

Bail et al.~\cite{2018Bail}
surveyed a large sample of Democrats and Republicans who visit Twitter at least three times a week about a range of social policy issues. 
One week later, they randomly assigned respondents to a treatment condition. They were offered financial incentives to follow a Twitter bot for one month. This bot exposed them to messages from those with opposing political ideologies. 
Respondents were resurveyed at the end of the month to measure the effect of this treatment, and at regular intervals throughout the study period to monitor treatment compliance. 
The authors found that Republicans who followed a liberal Twitter bot became substantially more conservative after treatment. 
Democrats exhibited small increases in liberal attitudes after following a conservative Twitter bot, although without statistical significance.
The authors found no evidence that exposing Twitter users to opposing views reduce the political polarization. 
The study indicated that attempts to introduce people to a broad range of opposing political views on a social network such as Twitter might be not only ineffective, but counterproductive. 

Gillani et al.~\cite{2018Gillani} 
sought to mitigate political echo chambers by showing the participants a subset of their social networks and asking them to discover their level of social connectivity. 
The authors created a web application. 
Each participant answered questions regarding their engagement in political discourse on Twitter. 
Next, the application presented a visualization of the participant’s network. 
The application then asked the participant to give their location in the network. 
After a guess was made, the tool revealed the true position of the participant. 
Sometimes the participant might also see a list of suggested accounts to follow that would increase their diversity score. 
Finally, the participant was asked to complete a post-survey.
For each participant, the authors measured the difference in their answers to the survey questions and the political diversity of the accounts this participant followed on Twitter before and after treatment. 
Participants asked to find their accounts in their social network, with nodes colored by inferred political ideology, tended to increase their belief in how ideologically closed they really were, but the political diversity of who they chose to follow actually decreased several weeks after treatment. 
The diversity of the followees of participants recommended to follow Twitter accounts with opposing political views increased one week after treatment. 

Chen et al.~\cite{2018Chen} 
relied purely on the topology of Friedkin-Johnsen’s model~\cite{1990Friedkin} of opinion formation to quantify the risk of conflict in a social network, 
A probabilistic model was proposed, assuming that the internal opinions of the $n$ nodes of the network follow a uniform distribution over $\{-1, 1\}^n$. 
The average-case conflict risk is defined as the expected conflict with regard to the internal opinions. 
An alternative and more robust measure, the worst-case conflict risk, is defined as the maximum conflict over all possible internal opinion vectors. 
They showed how both risk measures can be minimized by locally editing the network for a number of pre-existing measures of conflict and disagreement. 
Two 
algorithms 
were proposed to locally edit the network to reduce the worst-case and average-case risks for a number of measures of conflict. 
The authors focused on identifying a limited number of edges to add or remove in the network to reduce the risk of conflict. 
They showed the usefulness of these characterizations of conflict risk in a range of networks and claimed that their optimization minimized the actual risk on some random opinion assignments. 

Interian et al.~\cite{2020IntMorRib,2021IntMorRib} 
proposed the minimum intervention principle, which assumes that the smallest number of changes should be made in the original network by
any polarization reduction method. 
The issue of the insufficient communication between the polarized groups is solved by edge additions. 
The minimum cardinality edge addition problem is proposed as a strategy for reducing the polarization in real-world networks. 
The 
problem was shown to be NP-hard. 
Preliminary results obtained by an iterated greedy heuristic were presented in~\cite{mic2019}, while
three integer programming formulations were 
compared in~\cite{2021IntMorRib} with computational results for both randomly generated and real-life instances. It was shown that the polarization could be reduced to the desired threshold by the addition of few edges, as established by the minimum intervention principle that guided the problem formulation. 
According to the authors, there is often a straightforward way of spreading polarization-breaking information, even in strongly polarized networks. 

Chitra and Musco~\cite{2020Chitra} 
augmented the Friedkin-Johnsen opinion dynamics model~\cite{1990Friedkin} to include the filter bubble, the practice of connecting users with ideas they are already likely to agree with.
A network administrator is introduced in the model as an external actor that dynamically adjusts the strength of specific edges of a social network. 
The network administrator seeks to minimize a standard measure of disagreement between interacting users in the social network, since the authors considered that user engagement would increase by reducing users' disagreement. 
Individuals update their opinions according to the model's dynamics, and the administrator repeatedly adjusts the underlying network graph to achieve its own goal. 
The study 
showed that in Reddit and Twitter networks, after introducing the network administrator dynamics, even small changes to the edge weights may significantly increase the polarization. 
Finally, a simple modification in the network administrator's incentives that limit the filter bubble effect 
was proposed for countering the increasing polarization.
According to the authors, their solution increased user disagreement from 3\% to only 5\%, showing that this modification would minimally affect user engagement.

Santos et al.~\cite{2021Santos} 
investigated the relationship between social networks and reputation-based cooperation in large populations, analyzing the impact of network topology on polarization. 
They proposed a game-theoretical evolutionary model and studied dynamics in networks with varying degrees of community structure. 
They showed that networks exhibiting modular structures hamper global cooperation. 
Strategies of cooperating exclusively with in-group members fixate, sustaining polarization and group bias. 
The model uses the stern-judging social norm:
helping a good individual or refusing help to a bad one leads to a good reputation, whereas refusing help to a good individual or helping a bad one leads to a bad reputation. 
When communities are well-defined and reputations are attributed following stern-judging, polarization and group bias emerge: cooperation thrives within communities, though not across communities. 
Global cooperation is recovered as long as inter-community edges are added. 

Haddadan et al.~\cite{2021Haddadan} 
argued that structural bias may trap a user of the World Wide Web in a polarized bubble with no access to diversified opinions. 
They modeled user behavior by random walks, and defined the polarized bubble radius (BR) of a node as a measure to quantify its polarity. 
It denotes the expected number of steps to go from this node to a page of a different opinion. 
A node is in a polarized bubble if the expected length of the random walk to a page of different opinion is large. 
The structural bias of the whole graph is the sum of the radii of its polarized bubbles. 
The authors studied the problem of decreasing the structural bias using edge insertions. 
As ``healing'' all nodes with a high polarized bubble radius is hard to approximate, 
they presented an algorithm for finding the best set with a fixed number of edges whose insertion maximally reduces the graph's structural bias. 
The algorithm is able to return 
a constant-factor approximation using a greedy approach based on a specific variant of the random-walk closeness centrality~\cite{2003White}. 

Balietti et al.~\cite{Balietti2021} used informal political discussions with individuals sharing personal characteristics and social context to decrease polarization by exposing them to personal messages about a divisive political topic. 
According to the authors, friendship networks exhibit greater diversity of political views than is apparent to their members, and incidental conversations may  
expose interlocutors to diverse viewpoints. 
A large-scale experiment is performed by matching participants to peers having common interests and demographics and exposing them to a personal message about wealth redistribution. 
As a result, informal 
communication increases support for redistribution and reduces opinion polarization, suggesting that incidental conversations have the effect of increasing consensus on divisive and partisan topics. 
The authors 
showed that ``feeling close to the match'' is associated with an increase in the probability of assimilation of diverse political views. 

\subsection{Node modifications}

The approaches presented in this section are based on the hypothesis that introducing specific types of nodes (informed agents~\cite{2019Ghezelbash}, zealots~\cite{2020Shekatkar}) or changing the features of some nodes may decrease the network polarization. 

Ghezelbash et al.~\cite{2019Ghezelbash}
investigated how a group of selected informed agents can lead society towards some desired opinion. 
Informed agents are individuals 
aware of the desired opinion and act as hidden advisers, leading the society to this opinion 
by interactions with other individuals. 
An optimization technique was developed to solve the informed agent selection problem. 
The opinion dynamics model uses a network modeled by a
connected graph. 
The opinion of each agent on a certain issue can be represented by a real value 
in the interval $[-1, 1]$.
Opinions at the two ends of the interval are extreme opinions. 
If the members of each group reach a consensus on some specific opinion, there is complete polarization. 
Fragmentation occurs when
the divergence of opinions leads to more than two groups. 
Several goals such as polarization, fragmentation, and diversity of opinions were considered and formulated, showing that they are NP-hard optimization problems. 
For a specific sample graph, they were solved to optimality. 
Although the agents with more connections have more influence on the network, the authors showed that they are not necessarily the best candidates for being informed agents. 

Shekatkar~\cite{2020Shekatkar} 
considered zealots, or ``inflexible minorities'', as nodes in a social network that do not change their opinions under social pressure, investigating their effect on the polarization dynamics. 
The author proposed a quantifier called ``correlated polarization'' to measure the amount of network polarization. 
The correlated polarization is close to one if two extreme groups of comparable sizes and with opposite opinions are formed. 
Two types of zealots are studied: uniform and topology-based (when only high degree vertices are zealots). 
The polarization dynamics is simulated by using a simple majority rule~\cite{2002Galam} model. 
Considering an undirected network, every model's node could be in three different states: +1, -1, or 0, where +1 and -1 represent the opposite opinions and 0 corresponds to the neutral point of view. 
Each node in the network can be a zealot with some probability.
If a zealot has an already definite opinion, it will never change its state. 
The results indicated that the presence of zealots in a social network does not have a fixed effect and can lead to either positive or negative changes in the polarization, depending on the initial conditions and other factors, such as the edge density. \\

\subsection{Network design or recommendation modifications}
\label{sec:network_design}

The publications in this section propose strategies for reducing polarization based on changing the very design of the social network. 

A number of studies stand that social networks are in themselves causally sufficient to promote echo chambers due to their structure~\cite{2018Musco}, size~\cite{2018Madsen}, lack of trust~\cite{2018Koidl}, or the embedded recommendation systems~\cite{2019Grossetti}. 
Therefore, they argue that promoting features such as transparency~\cite{2019Mendes}, trust~\cite{2018Koidl}, or improved recommendation systems~\cite{2019Grossetti} can reduce polarization. 

In particular, some studies indicate that content personalization produced by recommendation systems may increase the echo chamber effect and create filter bubbles~\cite{2019Grossetti}. Therefore, some authors claim that modifying recommendation systems may reduce network polarization~\cite{2017Antikacioglu,2019Grossetti,2021RamaciottiMorales}. 

O'Hara and Stevens~\cite{2015OHara}
considered whether 
filtering and recommendation technology on the Internet could amplify groups' viewpoints, leading to polarization of opinion across communities and increases in extremism. 
The echo chamber arguments of Sunstein~\cite{2007Sunstein} were taken as representatives of this point of view and examined in detail in the context of a range of research in political science and the sociology of religion. 
The question was not whether there are echo chambers on the Internet; for there is plenty of evidence that there are. The two key
questions were therefore whether the Internet is complicit in the growth of echo
chambers, and if so, whether it should be the target of a policy focus.
The authors claimed that the support for the echo chamber thesis is not strong enough to justify regulation or intervention. 

Cremonini and Casamassima~\cite{2017Cremonini}
studied control strategies for social networks based on the introduction of random information into some selected driver agents. 
The goal was to better distribute knowledge among the agents by reducing polarization and augmenting their average skill level. 
The authors defined two information diffusion metrics. 
The average knowledge is the average skill with respect to the topics actually known by the agents. 
The knowledge diffusion, instead, is the average skill of agents with respect to the total number of topics in the network. 
The network tends to polarize when the difference between the two metrics increases, because the agents are gaining skills mostly on the same subset of topics without a corresponding increase in erudition. 
The authors studied how the control strategies affected the diffusion of topics and skills in the agent population.
The two strategies they studied were based, first, on the selection of a few influencers and the manipulation of their behavior, and, second, on the selection of many ordinary users as the drivers of the network. 
They concluded that the strategic use of random information could represent a realistic approach to network controllability and that both strategies could achieve this control effect.

Antikacioglu and Ravi~\cite{2017Antikacioglu} 
affirmed that collaborative filtering is a powerful framework for building recommendation systems. 
The propensity of such systems to favor popular products and create echo chambers has been observed. 
The authors addressed the problem of increasing diversity in recommendation systems based on collaborative filtering that use past ratings to predict a rating quality for potential recommendations. 
They formulated recommendation system design as a subgraph selection problem from a candidate super-graph of potential recommendations where both diversity and rating quality are explicitly optimized. On the modeling side, they defined a new flexible notion of diversity that allows a system designer to prescribe the number of recommendations each item should receive, and smoothly penalizes deviations from this distribution. On the algorithmic side, they showed that minimum-cost network flow methods yield fast algorithms for designing recommendation subgraphs that optimize this notion of diversity. They claimed the effectiveness of their model and method to increase diversity while maintaining high rating quality with empirical results in standard rating data sets from Netflix and MovieLens. 

Koidl~\cite{2018Koidl} 
proposed a novel design of social media applications, whose main motivation is the creation of a social network architecture that follows a ``trust by design'' paradigm. 
The author argued that the concept of trust is the driving force behind all three most important challenges of existing social networks, i.e., the existence of a filter bubble, the spreading of fake news, and the growing of echo chambers. 
For each user, a decentralized network is created. 
Due to the dynamics within the user environment, 
the author suggested a cloud-based storage solution. 
Each decentralized network follows a peer-to-peer architecture, in order to ensure a high level of privacy and control. 
The application's main feature is to establish computational trust by enabling a numerical trust value towards each new element within a social network, empowering the user to assess what elements to trust. 
To compute trust, different approaches were proposed. 
The result is a trust graph that informs the social graph about the trustworthiness of each element in the network. 
The authors claimed that this trust-based platform limits the ability to create fake or distorted representations of the individual, and strongly relies on authenticity, excluding fake or unauthentic representations. 

Madsen et al.~\cite{2018Madsen} 
employed an agent-based model that allows for relevant cognitive functions (Bayesian belief revision) and agent interaction (sharing their beliefs) to explore the emergent echo chambers. 
They showed that echo chambers can emerge among error-free Bayesian agents, and that larger networks encourage rather than ameliorate the growth of echo chambers. 
The authors tested interventions to reduce the formation of echo chambers, finding that system-wide truthful ``educational'' broadcasts ameliorate the effect, but do not remove it entirely. 
Such interventions are shown to be more effective on agents newer to the network. 
The authors claimed that social networks are in themselves causally sufficient to promote echo chambers. 
This carries a critical implication for interventions aimed at reducing them: individual-based interventions may help reduce somewhat the harmful or erroneous thinking that promotes the formation of echo chambers, but are not sufficient to remove them. 
Instead, the authors argued that system-based interventions might be more effective, 
taking advantage of top-down system alterations for reducing echo chambers. 

Musco et al.~\cite{2018Musco} 
explored the trade-off between disagreement and polarization in online social networks. 
Their research question was: given $n$ agents, each with its own initial opinion on a topic, and an opinion dynamics model, what is the structure of a social network that minimizes disagreement and controversy simultaneously? 
This question is central to recommendation systems: should a recommendation system prefer a link suggestion between two users with similar mindsets (to keep disagreement low), or between two users with different opinions to expose each other to a contrarian viewpoint (decreasing overall levels of polarization)? 
The authors provided a mathematical formulation for finding an optimal topology that minimizes the sum of polarization and disagreement under the Friedkin-Johnsen opinion formation model~\cite{1990Friedkin}, which takes into account both consensus and disagreement in the opinion update process. 
They proved that there always exists a graph with $O(n / \epsilon^2)$ edges that provides a $(1 + \epsilon)$ approximation algorithm to the optimal solution. 
They performed an empirical study of their methods on synthetic and two real-world datasets (Twitter and Reddit),
finding that there is space to reduce both controversy and disagreement in real-world networks. 

Grossetti et al.~\cite{2019Grossetti,2021Grossetti} 
studied communities on a large Twitter dataset to quantify how recommendation systems affect users’ behavior, and how content personalization can increase the echo chamber effect and create filter bubbles. 
A preliminary study was conducted to detect a filter bubble effect on users' community profiles, proposing a community-aware model 
whose objective is to reduce the filter-bubble impact. 
This model can be deployed on top of any existing recommendation system, enhancing it with a new scoring function that permits re-ranking the recommendations. 
To determine the similarity between communities, the authors considered (1) topology, (2) semantic information, and (3) flows of information between these communities. 
The model works with a set of recommendations, selected from the recommendation list produced by the recommendation system, and a community score vector that matches as much as possible the user profile. 
The authors showed that their solution 
improved the quality of recommendations by matching more closely the users' community profile and by reducing the filter bubble effect at a limited computation cost. 

Mendes et al.~\cite{2019Mendes} 
presented a platform for crowdsourced social participation that increases engagement and counteracts the formation of opinion bubbles and the echo chamber effect of social networks. 
The authors argued that 
clearly separated opinion groups could not necessarily indicate polarization, but might instead stem from rational disagreements. 
Polarized discussions arise from distorted perceptions about the other group’s motivations and points of view. 
The platform organizes topics of discussion around ``conversations''. 
Users can insert comments associated with a conversation or vote if they agree or disagree with other user's propositions. 
As the users interact with some conversation, it gradually becomes possible to recognize the different opinion profiles. 
As soon as the platform can classify users into distinct opinion groups, it displays results to all participants. 
Therefore, not only the owners of each conversation can extract useful metrics to guide policy and decision making. 
Each participant can access its own opinion profile or those from their network of peers and compare it with the whole. 
According to the authors, the proposed form of transparency towards information is an important factor to counteract the echo chamber effect. 

Sacharidis~\cite{2019Sacharidis} 
studied the effect of social-based recommendation systems in the diversity and novelty of the recommendations they make, questioning whether they lead to the formation of echo chambers. 
Social-based recommendation systems seek to exploit the effects of homophily and influence observed in social networks in order to improve their accuracy. 
This is achieved by enforcing similar preferences among users that are socially connected. 
The Douban dataset used 
in the numerical evaluation concerns a popular Chinese social networking service that allows users to connect to each other and provide content and ratings to movies, books, music, and events. 
The results indicated that social-based recommendations can often increase the diversity and novelty of user recommendations when measured individually and when examined with respect to the social groups to which users belong. 
The author also claimed that the social-based recommendations resulted in more accurate recommendations, while not sacrificing diversity and novelty.

Donkers and Ziegler~\cite{2021Donkers} 
alleged that, while most scientific work has framed echo chambers due to imbalances between polarized communities, members of the echo chambers often actively discredit outside sources to maintain coherent world views. 
They argued that two different types of echo chambers occur in social media: epistemic echo chambers create information gaps mainly through their structure, whereas ideological 
ones
systematically exclude counter-attitudinal information. 
An agent-based modeling approach was applied to investigate the characteristics of this dual echo chamber view. 
To assess the depolarizing effects of diversified recommendations, the authors relied on knowledge graph embeddings~\cite{2020Dai}. 
For community detection, they employed the Louvain algorithm~\cite{2011DeMeo}. 
To quantify the segregation between communities, modularity and homophily measures were used. 
The results showed that counteracting the two different types of echo chambers may require fundamentally different diversification strategies. 
Moreover, interventions seem to be most effective when a discussion first emerges. 
This shows the importance of what people observe when they first engage with a new topic. 

Morales and Cointet~\cite{2021RamaciottiMorales} 
studied the impact of recommendation systems on polarization. 
They presented the analysis of friend recommendations using real-world networks, where nodes (users) have dynamical positions in a ideological space, and where dimensions are indicators of attitudes towards issues in the public debate. 
The network evolves following a recommendation system, and the users opinions co-evolve following a DeGroot opinion model~\cite{1974DeGroot}. 
The authors applied the Duclos-Esteban-Ray  polarization measure~\cite{2004Duclos}, which is an extension of the Esteban-Ray measure~\cite{1994Esteban} of the distribution of user attitudes in the ideological space. 
The authors showed that different well-known recommendation systems can modify the convergence or divergence of social systems, affecting the evolution of polarization. 
For evaluating the effects of different recommendation systems on polarization, the authors used subgraphs of the Twitter network in the vicinity of French parliamentarians. 
The results indicated that some recommendation systems can decrease polarization, 
while others can increase it,  
leading the authors to argue that the use of a specific recommendation system can drive or mitigate the polarization appearing in real social networks. 

Fabbri et al.~\cite{Fabbri2022} studied the problem of mitigating radicalization pathways that 
occur when a user is exposed to polarized content, subsequently receiving increasingly radicalized recommendations. 
The authors model the set of recommendations of a ``what-to-watch-next'' recommendation system as a $d$-regular directed graph where nodes correspond to content items, links to recommendations, and paths to 
user sessions. 
They measured the polarity of a node as the expected length of a random walk from that node to any node representing non-radicalized content. High segregation scores are associated to larger chances to get users trapped in radicalization pathways. 
The problem of reducing the prevalence of radicalization pathways consists of selecting a small number of edges to ``rewire'' (following a similar idea to that in ~\cite{2020IntMorRib,2021IntMorRib}) so the maximum segregation score among all radicalized nodes is minimized while maintaining the relevance of the recommendations. 
Finding the optimal set of recommendations to rewire is proved to be NP-hard 
and a greedy algorithm 
is proposed for its solution. 

\subsection{Other approaches}

This section discusses other polarization mitigation models and strategies that do not fit in the approaches exposed in the previous sections.

Chen and Rácz~\cite{2021Chen} 
affirmed that while some actors spread misinformation to push a specific agenda, others aim to disrupt the network by increasing disagreement and polarization, thereby destabilizing society. 
Motivated by this phenomenon, the authors introduced a simple adversarial model of network disruption, where an adversary can take over a limited number of user profiles in a social network with the aim of maximizing disagreement, 
defined as a measure of how much neighboring nodes disagree in their opinions across the network. 
Polarization was defined as the variance of the opinions of all nodes, multiplied by the number of nodes. 
The authors investigated their model both theoretically and empirically, showing that the adversary will always change the opinion of a taken-over profile to an extreme to maximize disruption. 
They also showed that the adversary can increase polarization at most linearly in the number of user profiles it takes over. 
An empirical study of six adversarial heuristics on synthetic and real-world Reddit and Twitter networks was presented. 
The key conclusion was that the adversary can significantly disrupt the network (increasing polarization) using simple methods, such as targeting centrists.  

Arruda et al.~\cite{2021Ferraz} 
approached the appearance of echo chambers under the so-called underdog effect, emphasizing the influence of contrarian opinions in a multi-opinion scenario. 
The underdog effect is the tendency to support the less popular option. 
A modified Sznajd opinion dynamics model~\cite{Benatti_2020} is used. 
The authors considered an adaptation of the Sznajd model with the possibility of friendship rewiring, performed on several network topologies. 
They analyzed the relationship between topology and opinion dynamics by considering two measures: opinion diversity and modularity. 
Two strategies were tested: (1) the agents can reconnect only with others sharing the same opinion; and (2) same as in the previous case, but with the agents reconnecting only within their limited neighborhood. 
The authors found that the underdog effect, if strong enough, can increase the heterogeneity of opinions. 
This effect decreases the possibility of echo chamber formation. 
The number of opinions did not strongly affect the steady-state of the network dynamics.

\section{Conclusions}
\label{sec:discussion}

This review examined the most used network polarization measures and polarization reduction strategies. 
The use of measures based on homophily, modularity, random walks, content qualification, and balance theory has been proposed for measuring network polarization, also called in some studies by the terms controversy,
disagreement, and
conflict.
Polarization reduction strategies included node or edge modifications (including edge insertions or deletions, and adjustments in edge weights), changes in network design, and changes in the recommendation systems embedded in the networks. 

These polarization measures and reduction strategies have been used
in many case studies and real-life applications in different fields, such as partisan and political polarization, polarization in digital media consumption, climate change discussions, vaccination debates, and scientific co-authorship and collaboration.

Some studies analyzed partisan polarization in parliaments, online participatory platforms, web, and blogging networks in the USA~\cite{2021Dinkelberg,2017Garimella_Long_term,2007Reese,2017Shi,2008Zhang}, Italy~\cite{2015DalMaso}, Switzerland~\cite{2015Garcia}, Germany~\cite{2019Markgraf}, France~\cite{2021RamaciottiMorales}, Israel~\cite{2021Wolfowicz}, Venezuela~\cite{2015Morales}, India and Pakistan~\cite{2020Haq}, and a group of 16 European countries~\cite{2010maoz}. 
For instance, Markgraf and Schoch~\cite{2019Markgraf} 
reported a case study based on data from the German Federal Election of 2017 for illustrating their echo chambers research framework. 
Morales and Cointet~\cite{2021RamaciottiMorales} 
used subgraphs of the Twitter network in the vicinity of French parliamentarians, reporting that some recommendation systems can decrease polarization, while others can increase it. 
Maoz~\cite{2006Maoz}, instead, analyzed the political polarization of alliances between different countries and its effect on conflicts among states. 

Other authors studied online news consumption in mass media, newspapers, social networks, and blogs to identify factors that lead to polarization. 
Flaxman et al.~\cite{2016Flaxman} 
examined the web-browsing patterns of 50,000 anonymized US-located Internet users, 
showing that articles found via social media or web-search engines are associated with higher ideological segregation than those found 
directly visiting news sites. 
Ksiazek~\cite{2011Ksiazek} and Webster and Ksiazek~\cite{2012Webster} studied audience fragmentation, investigating how people allocate their attention across digital media. The authors found little evidence that audiences were composed of devoted loyalists, but the results suggested the presence of linguistic polarization. 
Tien et al.~\cite{2020Tien} 
compared groups of Twitter users who participated in the conversation about the Charlottesville events~\cite{misc_charlottesville} in the USA, finding that the retweet network largely splits according to user media preferences. 

Climate change emerged as one of the most polarizing discussion topics. 
Cook and Lewandowsky~\cite{2016Cook} investigated belief polarization, showing that the worldview has a dominant influence on climate beliefs (the authors used free-market support as a proxy for participants' worldview). 
Jasny et al.~\cite{2018Jasny} investigated the information diffusion process among climate policy networks in the US, finding an increase in the number of arcs that generate echo chambers from 2010 to 2016. 
Samantray and Pin~\cite{2019Samantray} studied a model of belief polarization in social networks that considers, in addition to the homophily (see Section~\ref{subsec:homophily}), the information credibility. They concluded that tweets expressing anti-climate change beliefs are largely not credible to the broader society. 

The COVID-19 pandemic attracted great attention to the echo chambers of the anti-vaccine community, which led to a decline in vaccination rates. 
Horne et al.~\cite{2015Horne} showed that it is possible to successfully counter people's anti-vaccination attitudes by making them appreciate the consequences of failing to vaccinate their children. 
Cossard et al.~\cite{2020Cossard} analyzed the Italian vaccination debate on Twitter. The two sides of the debate, one formed by vaccine advocates and the other formed by users skeptical about vaccination, tended to ignore each other's content, potentially leaving skeptics' concerns unanswered.

Some authors evaluated the polarization in co-authorship and collaboration networks. 
Soós and Kampis~\cite{2011Soos} analyzed the leading Hungarian research organizations, comparing the diversity of their publication performance and the polarity of each researcher's profile. 
Leifeld~\cite{2018Leifeld} analyzed two research traditions (the hermeneutic 
and the nomological) 
in the co-authorship network of researchers in Germany and Switzerland. A higher similarity between researchers leads to a greater probability of co-authorship, showing a homophilic behaviour between hermeneutic and nomological researchers observed by philosophers of science. 


The anonymous communication provided by social networks, to a great extent protected by the distance between those who participate in the dialogue, creates an incentive to expose more extreme opinions without the usual constraints that characterize the behavior observed in physical or face-to-face interactions.
Individuals often soften their ideas during face-to-face interactions not to hurt other persons' sensibilities. 
The digital environment frees individuals to express their views more openly, without concerns about the possible reactions that these opinions may have on others, 
often triggering heated and polarizing attitudes. 
Attempts to encourage dialogue and improve inter-group communication might mitigate the extreme polarization in social networks. 

\section*{Acknowledgments}
The authors are grateful to the suggestions and recommendations of two anonymous referees who contributed to improve the readability of this article. Work of Ruben Interian was supported by FAPERJ post-doctorate research grant E-26/204.200/2021.
Work of Ruslán G. Marzo was supported by FAPERJ doctorate scholarship E-26/200.330/2020.
Isela Mendoza was supported by CNPq doctorate scholarship 143289/2021-7.
Celso C. Ribeiro was supported by research grants CNPq 309869/2020-0 and FAPERJ E-26/200.926/2021. 
This work was also partially sponsored by Coordena\c{c}\~ao de Aperfeiçoamento de Pessoal de N\'\i vel Superior (CAPES), under Finance Code 001.

\bibliographystyle{plain}
\bibliography{main}

\end{document}